\documentclass[aps,preprint]{revtex4}%
\usepackage{amsfonts}
\usepackage{amsmath}
\usepackage{amssymb}
\usepackage{graphicx}%
\providecommand{\U}[1]{\protect\rule{.1in}{.1in}}
\begin{document}
\preprint{ }
\title[ ]{Singular inverse-square potential: renormalization and self-adjoint extensions
for medium to weak coupling}
\author{Djamil Bouaziz}
\email{djamilbouaziz@univ-jijel.dz}
\affiliation{Laboratoire de Physique Th\'{e}orique, Universit\'{e} de Jijel, BP 98, Ouled
Aissa, 18000 Jijel, Algeria}
\author{Michel Bawin }
\email{michelbw@skynet.be}
\affiliation{Institut de Physique B5, Universit\'{e} de Li\`{e}ge,  Sart Tilman 4000
Li\`{e}ge 1, Belgium\footnote{Present address: 8, Tige Paquette, 4550 Nandrin,
Belgium.}.}
\keywords{one two three}
\pacs{PACS number(s) 03.65.Ge, 03.65.Ca, 34.20.-b, 11.10.Hi}

\begin{abstract}
We study the radial Schr\"{o}dinger equation for a particle of mass $m$ in the
field of the inverse-square potential $\alpha/r^{2}$ in the
medium-weak-coupling region, i.e., with $-1/4\leq2m\alpha/\hbar^{2}\leq3/4$.
By using the renormalization method of Beane \textit{et} \textit{al., }with
two regularization potentials, a spherical square well and a spherical
$\delta$ shell, we illustrate that the procedure of renormalization is
independent of the choice of the regularization counterterm. We show that, in
the aforementioned range of the coupling constant $\alpha$, there exists at
most one bound state, in complete agreement with the method of self-adjoint
extensions. We explicitly show that this bound state is due to the attractive
square-well and delta-function counterterms present in the renormalization
scheme. Our result for $2m\alpha/\hbar^{2}=-1/4$ is in contradiction with some
results in the literature.

\end{abstract}
\volumeyear{2014}
\volumenumber{number}
\issuenumber{number}
\eid{identifier}
\date[Date text]{date}
\received[Received text]{date}

\revised[Revised text]{date}

\accepted[Accepted text]{date}

\published[Published text]{date}

\startpage{1}
\maketitle

\section{INTRODUCTION}

Once considered devoid of physical interest \cite{1}, the nonrelativistic
attractive singular inverse-square interaction was recently studied in
connection with many problems of great physical interest such as Efimov states
\cite{1.1}, dipole-bound anions in polar molecules \cite{1.2}, capture of
matter by black holes \cite{12}, atoms interacting with a charged wire
\cite{1.3}, and the dynamics of a dipole in a cosmic string background
\cite{1.4}. From a more formal viewpoint, the singular inverse-square
interaction provides a simple example of the renormalization-group limit cycle
in nonrelativistic quantum mechanics \cite{3,4,5,6}. It has been furthermore
shown \cite{4} that the method of self-adjoint extensions \cite{7,8} for
solving the Schr\"{o}dinger equation for a particle of mass $m$ with a strong
attractive inverse-square potential $V(r)=\alpha/r^{2}$ ($2m\alpha/\hbar
^{2}<-1/4$) is equivalent to the renormalization method (R-method) of Beane
\textit{et al}. \cite{3}. However, the method of self-adjoint extensions also
applies to the \textquotedblleft medium-weak\textquotedblright-coupling range
($-1/4\leq2m\alpha/\hbar^{2}\leq3/4$) and leads to the possible existence of a
single bound state even for a repulsive ($\alpha>0$) potential \cite{8,9}. It
has been argued \cite{10} that this counterintuitive result is an undesirable
feature of this method; the authors of Ref. \cite{10} discuss an alternative
regularization which removes this bound state. This problem is of physical
importance in the study of bound anions, where no bound states have been
experimentally observed for values of the strength of the dipole field below
some critical value (for a review, see Ref. \cite{11} ) and may be relevant to
the study of the near horizon structure of black holes \cite{12}. For the sake
of completeness, let us mention that the singular $\alpha/r^{2}$ potential was
studied in the framework of quantum mechanics with a minimal length
uncertainty relation \cite{1.5}. In this framework, the potential is
regularized in a natural way and the bound states only exist in the
strong-coupling regime.

The purpose of this paper is to apply the R-method of Ref. \cite{3} to study
the singular inverse square potential in the medium-weak-coupling range.
Following Ref. \cite{5}, we show, using square-well and delta-function
counterterms that, as in the strong-coupling regime, the R-method is
equivalent to the self-adjoint extensions technique. The R-method clearly
shows that a bound state may arise because the strength of the regularization
potential is such that, for a given choice of the self-adjoint extension, it
leads to a bound state even in the absence of long range inverse-square
interaction ($\alpha=0$) or in the presence of a repulsive ($0<2m\alpha
/\hbar^{2}\leq3/4$) long-range interaction. The R-method thus elucidates the
origin of this \textquotedblleft obscure\textquotedblright\ \cite{8} bound
state, which can, however, always be suppressed by a suitable choice of the
self-adjoint extension. In the special case $2m\alpha/\hbar^{2}=-1/4$, which
may be relevant to the near horizon structure of black holes \cite{12}, the
R-method predicts the existence of at most one bound state, in agreement with
Ref. \cite{9} but in sharp contrast with Refs. \cite{12} where it was claimed
that there are infinitely many bound states.

The rest of this paper is organized as follows. In Sec. II, we consider the
square-well regularization potential and study the renormalization of the
$\alpha/r^{2}$ interaction in the medium-weak-coupling range. In Sec. III we
renormalize the $\alpha/r^{2}$ potential using the spherical $\delta$ shell
regularization potential. We summarize our results in a brief concluding section.

\section{\bigskip SPHERICAL SQUARE-WELL REGULARIZATION}

The starting point of our study is the $s$-wave reduced radial Schr\"{o}dinger
equation for one particle of mass $m$ in the external potential $V\left(
r\right)  ,$%
\begin{equation}
\left(  -\frac{\hbar^{2}}{2m}\frac{d^{2}}{dr^{2}}+V\left(  r\right)  \right)
u\left(  r\right)  =Eu(r),\label{1}%
\end{equation}
where $V\left(  r\right)  $ is given by
\begin{align}
V\left(  r\right)   &  =\frac{\alpha}{r^{2}},\ \text{
\ \ \ \ \ \ \ \ \ \ \ \ \ (}r>R\text{),}\nonumber\\
&  =-\lambda\frac{\hbar^{2}}{2mR^{2}},\text{ \ \ \ (}r<R\text{).}\label{2}%
\end{align}
The dimensionless \textquotedblright coupling constant\textquotedblright%
\ $\lambda$\ is a positive function of the short-distance cutoff $R$ and the
long-range coupling constant $\alpha$ is taken such as $\left(  -1/4\leq
2m\alpha/\hbar^{2}\leq3/4\right)  .$

While the renormalization of the strong attractive $\alpha/r^{2}$ potential
($2m\alpha/\hbar^{2}<-1/4$) was studied extensively in the literature (see,
e.g., Refs. \cite{3,4,5,6}), the weak-medium-coupling regime $\left(
-1/4\leq2m\alpha/\hbar^{2}\leq3/4\right)  $ is studied here.

Following Refs. \cite{3,4,5}, we first consider Eq. (\ref{1}) in the zero
energy case ($E=0$). The solution that is continuous at $r=R$ is given by%

\begin{align}
u_{0}(r) &  =A\left[  \left(  r/r_{0}\right)  ^{1/2+\nu}-c\left(
r/r_{0}\right)  ^{1/2-\nu}\right]  ,\text{ \ \ \ }r>R\nonumber\\
&  =\frac{A}{\sin\sqrt{\lambda}}\left[  \left(  R/r_{0}\right)  ^{1/2+\nu
}-c\left(  R/r_{0}\right)  ^{1/2-\nu}\right]  \sin(k_{0}r),\text{\ \ \ \ }%
r<R,\label{3}%
\end{align}
where $k_{0}^{2}=\lambda/R^{2},$ $\nu=(2m\alpha/\hbar^{2}+1/4)^{1/2},$ $A$ is
a normalization constant, $c$ is an arbitrary constant, and\ $r_{0}$ is an
arbitrary scale. The usual matching condition of the derivative $u^{\prime
}(r)$ at $r=R$ yields%

\begin{equation}
\text{\ \ \ \ \ \ }\sqrt{\lambda}\cot\sqrt{\lambda}=\frac{\left(  \frac{1}%
{2}+\nu\right)  \left(  R/r_{0}\right)  ^{2\nu}-c\left(  \frac{1}{2}%
-\nu\right)  }{\left(  R/r_{0}\right)  ^{2\nu\ }-c}.\text{\ \ \ \ \ \ }
\label{4}%
\end{equation}
Equation (\ref{4}) gives the values of the short-range coupling constant
$\lambda$\ as a function of the cutoff $R.$

Now we turn to the bound-state spectrum ($E=-\frac{\hbar^{2}}{2m}k^{2}$). The
solution to Eqs (\ref{1}) and (\ref{2}) is then given by
\begin{align}
u(r) &  =Nr^{1/2}K_{\nu}(kr),\text{ \ \ \ \ \ \ }r>R,\nonumber\\
&  =N^{\prime}\sin(Kr),\text{ \ \ \ \ \ \ \ \ \ \ }r<R,\text{\ }\label{5}%
\end{align}
where $N$ and $N^{\prime}$ are normalization constants, $K_{\nu}(z)$ is a
modified Bessel function and $\ $%
\begin{equation}
K^{2}=k_{0}^{2}-k^{2}.\label{6}%
\end{equation}

Matching logarithmic derivatives at $r=R$ now gives%
\begin{equation}
KR\cot(KR)=\frac{1}{2}+kR\frac{K_{\nu}^{^{\prime}}(kR)}{K_{\nu}(kR)}.
\label{7}%
\end{equation}

In the low-energy limit $(kR\ll1),$ one has $KR\simeq k_{0}R=\sqrt{\lambda}.$
Using the formula \cite{13}%
\begin{align}
K_{\nu}(z) &  =\frac{\pi}{2}\frac{I_{-\nu}(z)-I_{\nu}(z)}{\sin\nu\pi
},\nonumber\\
I_{\nu}(z) &  \sim(\frac{1}{2}z)^{\nu}/\Gamma(\nu+1),\text{ \ \ }%
z\rightarrow0,\label{88}%
\end{align}
Eq. (\ref{7}) takes the following form in the limit $(kR\ll1)$:%
\begin{equation}
\sqrt{\lambda}\cot\sqrt{\lambda}=\frac{1}{2}+\nu\frac{\frac{\Gamma(1-\nu
)}{\Gamma(1+\nu)}\left(  \frac{kR}{2}\right)  ^{2\nu}+1}{\frac{\Gamma(1-\nu
)}{\Gamma(1+\nu)}\left(  \frac{kR}{2}\right)  ^{2\nu}-1}.\label{9}%
\end{equation}

By identifying Eqs. (\ref{4}) and (\ref{9}) we then get%

\begin{equation}
k=\frac{2}{r_{0}}\left[  \frac{\Gamma(1+\nu)}{c\Gamma(1-\nu)}\right]
^{1/2\nu}, \label{10}%
\end{equation}
with binding energy $E=-\hbar^{2}k^{2}/2m$. Note that the condition $kR\ll1$
implies that formula (\ref{10}) only holds for $R/r_{0}\ll1$.

Formula (\ref{10}) is in complete agreement with the result of Ref. \cite{8}
using the method of self-adjoint extensions. As one can easily see, in the
range $\left(  -1/4\leq2m\alpha/\hbar^{2}\leq3/4\right)  $ of the coupling
constant $\alpha$, there exists a single bound state even if the
inverse-square interaction is zero ($\nu=1/2$) or repulsive ($1/2<\nu\leq1$).
The origin of this bound state lies in the short-range modification of the
interaction (this will be clearer in the next section). On the other hand,
formula (\ref{10}) shows that this bound state can always be eliminated by
choosing the arbitrary constant $c<0$ and thus is not a compulsory feature of
the self-adjoint extension method.

Let us now examine the case $\nu=0$ (the coupling constant $\alpha$ takes the
critical value $-\hbar^{2}/8m$), which requires special consideration. Let us
recall that the inverse-square potential with this particular value of
$\alpha$ would have some physical importance as it appeared in the study of
the near horizon structure of black holes \cite{12}.

Denoting $v_{0}(r)$\ the zero energy solution for $\nu=0,$ we now have%
\begin{align}
v_{0}(r) &  =A\left(  r/r_{0}\right)  ^{1/2}\left[  1+c\ln\left(
r/r_{0}\right)  \right]  ,\text{ \ \ \ }r>R,\nonumber\\
&  =\frac{A}{\sin\sqrt{\lambda}}\left(  R/r_{0}\right)  ^{1/2}\left[
1+c\ln\left(  R/r_{0}\right)  \right]  \sin(k_{0}r),\text{\ \ \ \ }%
r<R.\label{11}%
\end{align}
The matching condition of the derivative $u^{\prime}(r)$ at $r=R$ now yields

\ \ \ \ \ \
\begin{equation}
\sqrt{\lambda}\cot\sqrt{\lambda}=\frac{1}{2}+\frac{c}{1+c\ln\left(
R/r_{0}\right)  }.\label{12}%
\end{equation}

The bound state wave function $v(r)$ is now%
\begin{align}
v(r)  &  =Nr^{1/2}K_{0}(kr),\text{ \ \ \ \ \ \ }r>R,\nonumber\\
&  =N^{\prime}\sin(Kr),\text{ \ \ \ \ \ \ \ \ \ \ }r<R,\text{\ } \label{13}%
\end{align}
where $N$ and $N^{\prime}$ are normalization constants, $k^{2}=-\frac
{2m}{\hbar^{2}}E$ and $K^{2}=k_{0}^{2}-k^{2}$ with $k_{0}^{2}=\lambda/R^{2}$.

As in the case $\nu\neq0$, the continuity of logarithmic derivatives at $r=R$
now gives the condition%
\begin{equation}
KR\cot(KR)=\frac{1}{2}+kR\frac{K_{0}^{^{\prime}}(kR)}{K_{0}(kR)}.\label{14}%
\end{equation}
For very shallow bound states, $kR\ll1$ $(KR\simeq k_{0}R=\sqrt{\lambda})$, we
use the formula \cite{13}%
\begin{equation}
K_{0}(z)\sim-\ln(z),\text{ \ \ \ }z\rightarrow0,\label{15}%
\end{equation}
and Eq. (\ref{14}) is written in the following form:%
\begin{equation}
\sqrt{\lambda}\cot\sqrt{\lambda}=\frac{1}{2}+\frac{1}{\ln(kR)}.\label{16}%
\end{equation}

From Eqs. (\ref{12}) and (\ref{16}), we get%

\begin{equation}
k=\frac{1}{r_{0}}\exp(1/c), \label{17}%
\end{equation}
with $R/r_{0}\ll1$. We obtain a single bound state with binding energy given
by \ %

\begin{equation}
E=-\frac{\hbar^{2}}{2mr_{0}^{2}}\exp(2/c).\label{18}%
\end{equation}
This result is in agreement with Ref. \cite{9} but is in strong disagreement
with some studies of the structure of near horizon black holes \cite{12}
advocating the existence of infinitely many bound states for $\nu=0$.

\section{REGULARIZATION WITH A $\delta$-SHELL POTENTIAL}

As it was noted in Ref. \cite{5}, renormalization theory should lead to
identical results for low-energy observables independently of the
regularization potential. It has been furthermore shown that the spherical
$\delta$ -shell potential is particularly convenient in the regularization of
the strong attractive inverse-square potential and gives the same results as
in the square-well regularization. In this section, we consider the
medium-weak-coupling range of the potential in the R-method with a spherical
$\delta$ shell potential. The conclusions of Sec. II will all be confirmed.

The potential that we consider now is given by%
\begin{align}
V\left(  r\right)   &  =\frac{\alpha}{r^{2}}\ \text{
\ \ \ \ \ \ \ \ \ \ \ \ \ \ \ \ \ \ \ \ \ }\left(  r>R\right)  \text{,}%
\nonumber\\
&  =-\frac{\lambda\hbar^{2}}{2mR}\delta(r-R^{-})\text{ \ \ \ }\left(
r<R\right)  \text{.}\label{19}%
\end{align}
with $\left(  -1/4\leq2m\alpha/\hbar^{2}\leq3/4\right)  $. The long-range
attractive $\alpha/r^{2}$ potential in Eq. (\ref{19}) is cut off at a short
distance $R$ by a spherical shell with radius $r=R^{-}$ infinitesimally close
to but smaller than $R$ \cite{5}.

As in the previous section, we first solve the zero energy Schr\"{o}dinger
equation in order to establish the renormalization-group flow. The solution to
Eq. (\ref{1}) with the potential (\ref{19}) that is continuous at $r=R$ in the
case $E=0$ is given by%

\begin{align}
u_{0}(r)  &  =A\left[  \left(  r/r_{0}\right)  ^{1/2+\nu}-c\left(
r/r_{0}\right)  ^{1/2-\nu}\right]  ,\text{
\ \ \ \ \ \ \ \ \ \ \ \ \ \ \ \ \ \ \ \ \ }r>R,\nonumber\\
&  =A\left[  \left(  R/r_{0}\right)  ^{-1/2+\nu}-c\left(  R/r_{0}\right)
^{-1/2-\nu}\right]  (r/r_{0}),\text{\ \ \ \ \ \ \ \ }r<R, \label{20}%
\end{align}
where $\nu=(2m\alpha/\hbar^{2}+1/4)^{1/2}$, $A$ is a normalization constant,
$c$ is an arbitrary constant and $r_{0}$ is an arbitrary scale.

The usual boundary condition from the delta-function potential at $r=R$ for
the derivative of the wave function is then \cite{5}
\begin{equation}
\lim_{r\longrightarrow R^{+}}r\frac{u_{0}^{\prime}(r)}{u_{0}(r)}%
-\lim_{r\longrightarrow R^{-}}r\frac{u_{0}^{\prime}(r)}{u_{0}(r)}=-\lambda(R).
\label{21}%
\end{equation}
From Eqs. (\ref{20}) and (\ref{21}) we then get%

\begin{equation}
\lambda(R)=\frac{\left(  1/2-\nu\right)  \left(  R/r_{0}\right)  ^{2\nu
}-c\left(  1/2+\nu\right)  }{\left(  R/r_{0}\right)  ^{2\nu}-c}\text{.}
\label{22}%
\end{equation}

We then proceed to the bound-state spectrum ($E=-\frac{\hbar^{2}k^{2}}{2m}$).
The radial wave function, the solution to Eqs. (\ref{1}) together with
(\ref{19}), has the form
\begin{align}
u(r) &  =Cr^{1/2}K_{\nu}(kr),\text{ \ \ \ \ \ \ \ \ \ }r>R,\nonumber\\
&  =C^{\prime}\sinh(Kr),\text{ \ \ \ \ \ \ \ \ \ \ }r<R,\text{\ }\label{23}%
\end{align}
where $C$ and $C^{\prime}$ are normalization constants. Using again the
boundary condition (\ref{21}), we find the following spectral equation:%
\begin{equation}
\frac{1}{2}+kR\frac{K_{\nu}^{^{\prime}}(kR)}{K_{\nu}(kR)}-kR\coth
(kR)=-\lambda(R).\label{24}%
\end{equation}

By using formulas (\ref{88}), we find from Eq. (\ref{24}) in the limit
$kR\ll1$,%
\begin{equation}
\lambda(R)=\frac{\nu+\frac{1}{2}+(\nu-\frac{1}{2})\frac{\Gamma(1-\nu)}%
{\Gamma(1+\nu)}\left(  \frac{kR}{2}\right)  ^{2\nu}}{1-\frac{\Gamma(1-\nu
)}{\Gamma(1+\nu)}\left(  \frac{kR}{2}\right)  ^{2\nu}}.\label{25}%
\end{equation}

From Eqs. (\ref{22}) and (\ref{25}) we then get%

\begin{equation}
k=\frac{2}{r_{0}}\left[  \frac{\Gamma(1+\nu)}{c\Gamma(1-\nu)}\right]
^{1/2\nu}. \label{26}%
\end{equation}

As it is expected, formula (\ref{26}) is exactly the same expression
(\ref{10}), obtained with a spherical square-well regularization potential,
and then it is in complete agreement with the method of self-adjoint
extensions \cite{8}. Recall that the arbitrary scale $r_{0}$, whereby
depending on the energy spectrum, must satisfy $R/r_{0}\ll1$ due to the
condition $kR\ll1$, and that the arbitrary constant $c$ must be positive.

As in the previous section, the occurrence of a bound state with formula
(\ref{26}) can be interpreted by the short-range modification of the
interaction. In fact, Eq. (\ref{22}) shows that the strength of the
counterterm may take positive values (thus leading to an attractive
delta-function potential) even if the inverse-square interaction is zero
($\nu=1/2$) or repulsive ($1/2<\nu\leq1$). Thus, the R-method elucidates the
origin of this \textquotedblleft counterintuitive\textquotedblright\ bound
state, which can, however, always be suppressed, as in the self-adjoint
extension method, by choosing the parameter $c<0$.

We now consider the particular case $\nu=0$ ($2m\alpha/\hbar^{2}=-1/4$)
following the same procedure as in Sec. II.

The zero energy radial wave function $v_{0}(r)$ now reads
\begin{align}
v_{0}(r) &  =A\left(  r/r_{0}\right)  ^{1/2}\left[  1+c\ln\left(
r/r_{0}\right)  \right]  ,\text{ \ \ \ \ \ \ \ \ \ \ \ \ \ \ \ \ }%
r>R,\nonumber\\
&  =A\left(  R/r_{0}\right)  ^{-1/2}\left[  1+c\ln\left(  R/r_{0}\right)
\right]  \left(  r/r_{0}\right)  ,\text{\ \ \ \ }r<R.\label{27}%
\end{align}
The boundary condition (\ref{21}) from the delta-function potential at $r=R$
now gives \ \
\begin{equation}
\lambda\left(  R\right)  =\frac{1}{2}-\frac{c}{1+c\ln\left(  R/r_{0}\right)
}.\label{28}%
\end{equation}

The radial wave function for the bound state $v(r)$ has the form%
\begin{align}
v(r)  &  =Cr^{1/2}K_{0}(kr),\text{
\ \ \ \ \ \ \ \ \ \ \ \ \ \ \ \ \ \ \ \ \ \ \ \ \ \ \ \ \ \ }r>R,\nonumber\\
&  =CR^{1/2}\frac{K_{0}(kR)}{\sinh(KR)}\sinh(kr),\text{ \ \ \ \ \ \ \ \ \ \ }%
r<R,\text{\ } \label{29}%
\end{align}
where $C$ is a normalization constant and $k^{2}=-\frac{2mE}{\hbar^{2}}$. \ 

Imposing the boundary condition (\ref{21})\ to the solution (\ref{29}), we now
get for $kR\ll1$,%
\begin{equation}
\lambda\left(  R\right)  =-\frac{1}{2}-\frac{1}{\ln\left(  kR\right)
}+kR\coth kR.\label{30}%
\end{equation}
From Eqs. (\ref{28}) and (\ref{30}), we then find%

\begin{equation}
k=\frac{1}{r_{0}}\exp(1/c),\label{31}%
\end{equation}
with $R/r_{0}\ll1$. We again get a single bound state with exactly the same
expression of the binding energy as in the spherical square-well
regularization potential [see Eq. (\ref{17})]. The latter result confirms the
equivalence between the R-method of Beane \textit{et} \textit{al. }\cite{3}
and the method of self-adjoint extensions \cite{8,9}. On the other hand, our
conclusion is in contradiction with the claim of the authors of Refs.
\cite{12} advocating the existence of infinitely many bound states for $\nu=0$.

\section{ SUMMARY AND CONCLUSION}

We have applied the renormalization method of Beane \textit{et} \textit{al.
}\cite{3}\textit{ }and studied the problem of the singular-inverse square
potential $\alpha/r^{2}$ in the medium-weak-coupling region, i.e., with
$-1/4\leq2m\alpha/\hbar^{2}\leq3/4$, by using two regularization potentials,
as in Ref. \textit{ }\cite{5} : a spherical square well and a spherical
$\delta$ shell. As it was expected, the results obtained were independent of
\ the short-range interaction at low energy. We have explicitly shown that, in
both of these cases, there exist at most one bound state in the aforementioned
range of the coupling constant $\alpha$, as predicted by the method of
self-adjoint extensions. The expression of the bound-state energy has been
derived, and we have explicitly shown that this bound state is due to the
attractive short-range regularization potential, which corresponds to a given
choice of the self-adjoint extension that leads to a bound state even in the
absence of a long-range inverse-square interaction ($\alpha=0$) or in the
presence of a repulsive ($0<2m\alpha/\hbar^{2}\leq3/4$) long-range
interaction. The particular case $2m\alpha/\hbar^{2}=-1/4$, which may be
relevant to the study of the near horizon structure of black holes \cite{12},
has been considered separately. Renormalization leads also to, at most, a
single bound state. The latter result is in conflict with the statement of the
authors of Refs. \cite{12}, who claimed the existence of infinitely many bound
states for this critical value of $\alpha$. On the other hand, our results are
in complete agreement with the method of self-adjoint extensions. This leads
us to confirm the conclusion of Ref. \textit{ }\cite{4}, namely, the
equivalence between the R-method of Beane $et$ \textit{al.,} and the method of
self-adjoint extensions in the study of singular potentials.

\begin{acknowledgments}
The work of D.B is supported by the Algerian Ministry of Higher Education and
Scientific Research under the PNR Project No. 8/u18/4327 and the CNEPRU
Project No. D017201600026.
\end{acknowledgments}

\end{document}